\documentclass{svmult}

\usepackage{mathptmx}       
\usepackage{helvet}         
\usepackage{courier}        
\usepackage{type1cm}        
\usepackage{makeidx}         
\usepackage{graphicx}        
\usepackage{multicol}        
\usepackage[bottom]{footmisc}

\usepackage{url}
\usepackage{amsmath}
\usepackage{amssymb}
\usepackage[normalem]{ulem} 
\usepackage{epsfig}
\usepackage[curve]{xypic}

\begin{document}

\newcommand{\simplex}[1]{\left\langle #1 \right\rangle}
\newcommand{\mathset}[1]{\left\{#1\right\}}
\newcommand{\mathpset}[2]{\left\{#1 \mid #2 \right\}}
\newcommand{\undefined}{{\uparrow}}

\title*{Signed Simplicial Decomposition and Overlay of $n$-D Polytope Complexes}
\author{Norbert Paul}
\institute{Norbert Paul \at Karlsruhe Institute of Technology~(KIT),
           Geodetic Institute,
           Englerstra{\ss}e~7,
           76131~Karlsruhe,
           Germany,
           \email{norbert.paul@kit.edu}}

\titlerunning{Polytope Complex Decomposition and Overlay}
\authorrunning{Norbert Paul}

\maketitle

\abstract{Polytope complexes are the generalisation of polygon meshes in geo-information 
          systems (GIS) to arbitrary dimension, and a natural concept for accessing spatio-temporal information. 
          Complexes of each dimension have a straight-forward dimension-independent database representation 
          called \emph{Relational Complex}. 
          Accordingly, complex overlay is the corresponding generalisation of map overlay in GIS to arbitrary 
          dimension. 
%
%
          Such overlay can be computed by partitioning the cells into simplices, intersecting these and finally 
          combine their intersections into the resulting overlay complex. 
          Simplex partitioning, however, can expensive in dimension higher than $2$.
          In the case of polytope complex overlay \emph{signed} simplicial decomposition is an alternative. 
          This paper presents a purely combinatoric polytope complex decomposition which ignores geometry. 
          In particular, this method is also a decomposition method for \emph{non-convex} polytopes. 
          Geometric $n$-D-simplex intersection is then done by a simplified active-set-method---a 
          well-known numerical optimisation method. 
          ``Summing'' up the simplex intersections then yields the desired overlay complex.}

\keywords{$n$-d spatial modelling, topology, geometry, overlay}

\section{Introduction}

An important query operation in GIS is the overlay of some given ``topologies'' to generate new such ``topologies''. 
An example could be cadastral land-owner data overlaid with environmental stress data which helps to identify which 
owner is affected by what averse environmental influences. 
This operation seems to be missing in 3D cadastral applications:
\begin{quotation}
  However, 3D data management and analysis such as querying,
  manipulation, 3D map overlay, 3D buffering have been largely neglected in spatial database
  systems and Geographic Information Systems. \cite{Streilein:3DDataManagement}
\end{quotation}
To compute such overlay by first triangulate the input, then overlay these triangles 
and finally recombine the resulting intersections into the desired overlay complex is possible in $2$-d, but  
such triangulation in $3$-d, however, has $\Omega(n^2)$ space complexity in general \cite{Chazelle:ConvexPartitions} 
and, hence, is very expensive. 

The special case of \emph{convex} $3$-d-shapes, however, is almost trivial: Fix an arbitrary interior point and make it 
the apex of a family of cones atop the boundary faces. If these faces have been triangulated before the result is a 
simplicial decomposition of the convex shape. 

Also the area of an arbitrary non-convex planar polygon can be computed by a signed sum of the triangle areas made up 
of one edge and a fixed arbitrary vertex in the polygon plane. 
These observations will here be generalised to arbitrary dimension and used to compute $n$-d complex intersection.

\section{Related Work}

Much work has been done on polytope decomposition, volume computation, and 
intersection---mostly, however, on convex polytopes represented by vertices (via convex hull) or by 
(intersecting) half spaces. 
That representations do not allow non-convex polytopes and some of the above mentioned problems 
(even vertex enumeration) are NP-hard if no fixed dimension upper bound is given
\cite{Khachiyan:hardPolyhedronVertices}. 
By using \emph{polytope complexes} instead, vertices, edges, faces etc.\ are already \emph{explicitly} 
enumerated as a precondition and the above problems can be avoided. 

Volume computation of convex polytopes is discussed in 
\cite{Lawrence:PolytopeVolume}
and in \cite{BuelerEtAl:VolumeComputation}
where the latter also introduces signed simplicial decomposition.

Most work on decomposition is dedicated to unsigned decomposition as studied, for example, in the 
survey \cite{Chazelle:DecompAlg}.
Unsigned simplicial decomposition of convex polytope complexes are known as 
Boundary Triangulation \cite{BuelerEtAl:VolumeComputation}, and as Cohen Hickey's Triangulation 
\cite{CohenHickey:Triangulation}, cited by \cite{BuelerEtAl:VolumeComputation}. 

A work very similar to this article is described in 
\cite{Bulbul:AHD} 
and in \cite{Bulbul:ConvexDecomp}
as ``alternate hierarchical decomposition'' (AHD). 
This is also a singed decomposition into convex parts and its authors, too, consider it useful to compute
intersection, union and symmetric difference between non-convex polytopes. 
However, with the shape
$$\begin{picture}(75,90)
\setlength{\unitlength}{1.5\unitlength}
\put(0,10){\line(2,1){20}}
\put(0,10){\line(5,4){25}}
\put(0,10){\line(0,1){40}}

\put(0,50){\line(5,-4){25}}
\put(0,50){\line(3,-1){30}}
\put(0,50){\line(2,1){20}}

\put(20,20){\line(1,2){5}}

\put(20,60){\line(3,-1){30}}

\put(25,30){\line(1,-6){5}}
\put(25,30){\line(1,2){5}}

\put(30,0){\line(2,1){20}}
\put(30,0){\line(0,1){40}}

\put(30,40){\line(2,1){20}}

\put(50,10){\line(0,1){40}}
\end{picture}$$
that method may not terminate. 

\section{Basic Notions and Data Model}

In many cases spatial data can be considered a model of some partitioning of the two- or three-dimensional space in 
spatial ``chunks'' like areas, volumes, faces etc.\ just like Computer-Aided Design (CAD), GIS, $3$-d city models, or 
subsoil geology models do. 

\begin{figure}
\begin{center}
\includegraphics[width=0.5\textwidth]{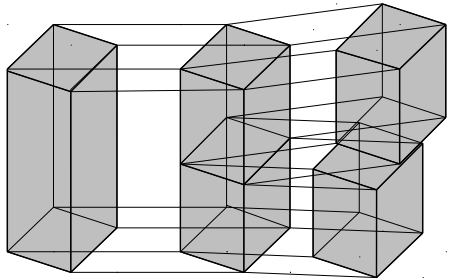}
\end{center}
\caption{The spatio-temporal process of a volume object (left) splitting into two (middle) and one of them 
slipping away over time (middle to right) as a $4$-d space-time-complex.}
\label{fig:faultslip}
\end{figure}

If such partitioning undergoes changes in time this can be considered a partitioning of the four-dimensional 
space-time into what might then be called ``hyperchunks''. A volume, for example, may extend over a time interval and then 
be split in two at a time point after which two such volumes start to exist. Then that volume at the interval before the split 
can be considered a four-dimensional ``hypervolume'' bounded by two volumes which mark the splitting event as shown in 
Figure~\ref{fig:faultslip}. 

\subsection{Topological Extension of the Relational Model}

What follows is a brief introduction into our topological data model \cite{BradleyPaul:DTop}.
Consider our aforementioned subdivision of space (or space-time) satisfying the following preconditions:
\begin{itemize}
\item A compact subset of the $n$-dimensional real space $\bbbr^n$ is subdivided into a finite number 
      of parts.
\item Each part is a connected $i$-dimensional manifold without boundary. 
\item Each such part is flat. That means it is within an $i$-dimensional affine subspace of $\bbbr^n$.
\item The boundary of each $i$-dimensional part is the union of parts of lower dimensions. 
\end{itemize}
We call this subdivision a \emph{finite polytope complex}.
Then the natural topology of the underlying $n$-dimensional real space $\bbbr^n$ generates 
a so-called \emph{quotient topology}. As the number of parts is finite, 
their topology is finite, too, and, hence, it is a so-called Alexandrov-topology \cite{Alexandroff:Raeume}
which has an important characteristic: It can be stored by a relation called ``incidence graph'' and so it fits 
into a relational database:
\begin{definition}[Topological Data Type]
Let $X$ be a set and $R\subseteq X\times X$ a relation on $X$. We call the pair $(X,R)$ a 
\emph{topological data type}. A subset $A \subseteq X$ is said to be \emph{open in $(X,R)$}, iff 
all $x\in X$ and $a \in A$ satisfy $x\,R\,a \Rightarrow x \in A$. 
The relation $R$ is also called the \emph{incidence relation} of $(X,R)$.
\end{definition}
So far we have only relabelled what is commonly known as simple directed graph. 
But note that every topology for a finite set can be stored in that simple manner. 

\begin{example}[Combinatorial Square]\label{exa:square1}
We partition a unit square $[0,1]\times[0,1]$ $\subset \mathbb{R}^2$ into nine elements: 
four vertices 
$a = \mathset{(0,1)}$,  
$b = \mathset{(1,1)}$,  
$c = \mathset{(0,0)}$,and 
$d = \mathset{(1,0)}$, 
four edges 
$e = {\left]0,1\right[}\times\mathset{1}$,
$f = \mathset{0}\times {\left]0,1\right[}$, 
$g = \mathset{1}\times {\left]0,1\right[}$, and
$h = {\left]0,1\right[}\times\mathset{0}$, 
and the face 
$F = {\left]0,1\right[}\times{\left]0,1\right[}$, which gives the topological space depicted 
at the left-hand side of the diagram below. 
At its right-hand side there is the corresponding topological datatype $(X,R)$ with point set 
$X=\mathset{a,b,c,d,e,f,g,h,F}$ and incidence relation 
$$
   R=\mathpset{(x,y)}{\text{The right image has an arrow $x\to y$.}}.
$$
$$
\xymatrix{
  \bullet 
  \ar@{-}[rr]^(-0.05)a^e^(1.05)b &   &\bullet    &&   a & e \ar[l]\ar[r]& b    
\\
                              & F &             &&    f \ar[u]\ar[d] & F \ar[u]\ar[d]\ar[r]\ar[l]
                                                                             & g\ar[u]\ar[d]
\\
\bullet 
\ar@{-}[uu]^f
\ar@{-}[rr]_(-0.05)c_h_(1.05)d &   &\bullet 
                                    \ar@{-}[uu]_g &&  c & h \ar[l]\ar[r]& d
}
$$
\end{example}
The justification for relabelling ``graph'' to ``topological data type'' is our adaption of continuous maps 
to topological data types: 
\begin{definition}[Continuous Database Map]
Let $(X,R)$ and $(Y,S)$ be topological data types and $f:X\to Y$ a map. 
$S^{*}$ denotes the transitive and reflexive closure of $S$. We then call $f$ a 
\emph{continuous database map} iff $(f(a),f(b))\in S^{*}$ holds for all $(a,b)\in R$. 
Then we write $f:(X,R)\to(Y,S)$.
\end{definition}
The continuous database maps are exactly the continuous maps between the corresponding Alexandrov spaces.
What we have in mind is a relational database table $X$ of spatial 
entities together with a table $R$ as an n:m relation type from $X$ to itself.

\subsection{Algebraic Topology in Relational Databases}

We now extend our above data model to algebraic topology:
A \emph{Relational Chain Complex} is a topological data type  $(X,R)$ if the relation 
$R$ also carries additional information about the orientation of each cell by attaching signs 
to the cell-cell-incidences. 

Here our partitioning of the space consists of a sequence $(C_n,\ldots,C_0)$ of sets $C_i$ of 
$i$-dimensional manifolds which together make up the entire space and which we call by abuse of 
language ``$i$-cells''. 
Note that the above preconditions guarantee, that at least our $0$-``cells'' and our $1$-``cells'' are 
cells indeed. 
Each $i+1$-cell $c$ is orientable and bounded by a set $D$ of $i$-cells. By fixing an orientation 
for each cell in our partitioning we specify a sign for each cell $d$ in $D$ which depends on $c$: 
It is positive if $c$ is touched by $d$'s ``front'' side, 
negative at its ``rear'' side, or zero if $d$ touches $c$ by ``both'' sides. 
We refer to \cite[p.\ 233]{Hatcher:AlgTop} for details on what ``front'' and ``rear'' in arbitrary dimension 
mean.

The above mentioned sign defines a function 
\begin{equation}
   \sigma_c:D\to \left\{-1,0,+1 \right\} ; d \mapsto \begin{cases}
                                                                + 1  &: \mbox{``front''}\\
                                                       \phantom{+}0  &: \mbox{``both''}\\
                                                                - 1  &: \mbox{``rear''}
                                                      \end{cases}\;.
\end{equation} 
Now we extend our relation $R$ from the topological data type $(X,R)$ to a function 
\begin{equation}
   M:R\to\bbbz,(x_{i+1},x_{i})\mapsto\sigma_{x_{i+i}}(x_i)
\end{equation}
and store it into a database table \texttt{M} with schema 
\begin{align*}
    &\texttt{M(\underline{cell}:$X,$ \underline{boundar}y:$X,$ sigma:$\bbbz$)}.
\\
\intertext{%
Considering \texttt{M} a sparse matrix the matrix product of \texttt{M} with itself 
can be computed by
}
&\begin{array}{ll}
\multicolumn{2}{l}{\tt create~view~M\_squared~as}\\
\tt select      &\tt~M1.cell,~M2.boundary,\\
                &\tt~sum(M1.sigma * M2.sigma)~as~sigma\\ 
\tt from        &\tt~M~M1,~M~M2\\
\tt where       &\tt~M1.boundary=M2.cell\\
\tt group~by    &\tt~M1.cell,~M2.boundary.
\end{array}
\\
\intertext{%
If \texttt{M\_squared} only contains zero entries it denotes a complex boundary. 
The above SQL-statement defines the multiplication 
of two sparse matrices \texttt{M1} and \texttt{M2} by simply replacing the 
\texttt{from}\nobreakdash-clause by
}
&\tt from~~M1,~M2.
\end{align*}

Now we can define our data model:
\begin{definition}[Relational Complex]
A sequence $(X_n,\ldots,X_0)$ of finite sets---the \emph{cells}---together 
with a sequence $(D_n,\ldots,D_1)$ of sparse $X_i\times X_{i-1}$-matrices---the 
\emph{boundaries}---is called a \emph{relational complex of dimension $n$} if every matrix 
product $D_{i+1}\cdot D_i$ of two consecutive matrices has only entries of value zero.
\end{definition}
This definition fixes a static dimension upper bound $n$.
\emph{Dynamic} dimension is also possible: Collect all cells into a table $X$, all matrices into a table 
$D$, and specify dimension merely by an integer attribute.
To specify geometry we simply store the $n$ coordinates of each 
vertex, say, in case $n=4$, by \texttt{x}, \texttt{y}, \texttt{z}, and \texttt{t}.

\begin{example}[Combinatorial Square]
The following relational complex for Example \ref{exa:square1} has 
Edges $e$ and $h$ running from left to right whereas Edges $f$ and $g$ point downwards. 
Face $F$ gets a counter-clockwise orientation:

The right-hand side shows the corresponding relational complex with point sets 
$X_0=\mathset{a,b,c,d}$, $X_1=\mathset{e,f,g,h}$, and $X_2=\mathset{F}$ and boundaries 
$$
   R_i=\mathpset{(x,y,\pm 1)}{x\in X_i, y \in X_{i-1}, \text{arrow $x \stackrel{\pm\,}{\longrightarrow} y$ in right image below.}}.
$$
$$
\xymatrix{
  \bullet 
  \ar@{->}[rr]^(-0.05)a^e^(1.05)b &   &\bullet       &   & a & e \ar[l]_{-}\ar[r]^{+}& b    
\\
                                  & F\ar@{}[ld]|(0.25){}="a"
                                     \ar@{}[rd]|(0.25){}="b"
                                     \ar@{}[ru]|(0.25){}="c"
                                     \ar@{.>}'"a";"b" '"c"
                                      &               &   & f \ar[u]^{-}
                                                              \ar[d]_{+} 
                                                              & F \ar[u]^{-}
                                                                  \ar[d]_{+}
                                                                  \ar[r]^{-}
                                                                  \ar[l]_{+}
                                                                             & g
                                                                               \ar[u]_{-}
                                                                               \ar[d]^{+}
\\
\bullet 
\ar@{<-}[uu]^f
\ar@{->}[rr]_(-0.05)c_h_(1.05)d &   &\bullet 
                                    \ar@{<-}[uu]_g &&  c & h \ar[l]^{-}\ar[r]_{+}& d
}
$$
Note that the orientation of Edge $h$ is ``compatible'' with the orientation of Face $F$ so 
the entry in the boundary table is the tuple $(F,h,+1)$, whereas $g$ runs contrary to the face 
orientation as entry $(F,h,-1)$ indicates. 

Now consider all paths from face $F$ to vertex $a$:
$$
\xymatrix@1{F\ar[r]^{-1}&e\ar[r]^{-1}&a & \text{and} & F \ar[r]^{+1}&f\ar[r]^{-1}&a}.
$$
When we multiply the coefficients of the first path we get $(-1)\times(-1) = +1$ 
and for the second path we have $(+1)\times(-1) = -1$. 
The sum of both is zero---the $F,a$-entry of the matrix product. 
\end{example}

We also need a mapping between two relational complexes:
\begin{sloppy}
\begin{definition}[Relational Complex Morphism]
Let $(X_n,\ldots,X_0)$ with boundaries $(D_n,\ldots,D_1)$ be a relational complex, and 
let $(Y_n,\ldots,Y_0)$ with boundaries $(B_n,\ldots,B_1)$ be another relational complex. 
Then we call a sequence $(F_n,\ldots,F_0)$ of (sparse) $X_i\times Y_i$-matrices 
a \emph{relational complex morphism}, if each $D_i\cdot F_{i-1} = F_i\cdot B_i$ except for zero entries. 
\end{definition}
\end{sloppy}

The above used difference $A-B$ of sparse matrices pads missing entries in one of the 
matrices with zero. From now on we say ``algebraically equal'' instead of 
``$a = b$ except for zero entries''. 

\section{Complex Overlay}

Now having presented our data model we introduce the problem:

Two polytope complexes $C$ and $K$ partitioning a compact part $|C|$ and 
$|K|$ of the real vector space into flat manifolds have a common refinement of the 
intersection $|C|\cap|K|$: the polytope complex of $|C|\cap|K|$ of the non-empty pair-wise intersections 
of cells in $C$ and $K$. We call this \emph{intersection} overlay and denote it by $C\cap K$.

Additionally there is a common refinement of $|C|\cup|K|$ which takes into account the exterior of the 
a complex if needed. This is \emph{union} overlay, denoted by $C\cup K$.

\begin{example}[Two Squares and a Triangle]
The union-overlay and intersection-overlay of a complex $S$ which consists of two meeting squares 
$ABDE$ and $BCEF$ with another complex $T$---simply the triangle $XYZ$---may be the 
following complexes $S\cup T$ (left) and $S\cap T$ (right):
\begin{align*}
\xymatrix{
&& X\ar@{-}[ddl]\ar@{-}[ddr]\\
A 
\ar@{-}[rr]
\ar@{}[rr]|<(0.72){\times} ^{AB\cap XY}
\ar@{-}[dd]
  &  & B
      \ar@{-}[rr]
      \ar@{}[rr]|<(0.175){\times}^{BC\cap XZ}
      \ar@{-}[dd]
      \ar@{}[dd]|{\times}
        &   & C 
              \ar@{-}[dd]                    \ar@{}[rr]|<(0.72){}="ABXY"
                                             \ar@{-}"ABXY";[rr]|<(0){\times}^<(0){AB\cap XY}
                                             &&    B
                                             \ar@{}[rr]|<(0.175){}="BCXZ"
                                             \ar@{-}"BCXZ"|<(1){\times}^<(1){BC\cap XZ}
                                             &&\\
  & Y 
    \ar@{-}[rr]_<(0.7){BE\cap YZ}
      &   & Z &                              &Y \ar@{-}"ABXY"
                                                & \times \ar@{-}[u]\ar@{-}[l]\ar@{-}[r]
                                                   & Z \ar@{-}"BCXZ" \\
 D 
 \ar@{-}[rr]  
   &   & E 
         \ar@{-}[rr]
           &   & F
}
\end{align*}
\end{example}

The problem is: How can the intersection complex be computed?

\section{Triangulation and Signed Decomposition}

Triangulation is the partitioning of polygons into triangles which practically is in $O(n \log n)$,
theoretically even in $O(n)$ \cite{Chazelle:Triangulation}.

The $3$-d-analog is partitioning a polyhedron into tetrahedra and is often called ``tetrahedralisation''. 
The worst case lower bound on the number of resulting tetrahedra is in $\Omega(n^2)$, 
where $n$ is the number of cells \cite{Chazelle:ConvexPartitions}.

The corresponding dimension-independent notion is ``simplicial decomposition''.
We know this is expensive but, happily, signed simplicial decomposition is often an alternative.

We will now present \emph{combinatorial} $n$-d signed simplicial decompositions which generalise 
the well-known methods Boundary Triangulation \cite{BuelerEtAl:VolumeComputation}, and 
Cohen Hickey's Triangulation (\cite{CohenHickey:Triangulation}, cited by \cite{BuelerEtAl:VolumeComputation})
by only using algebraic and topological information provided as relational complex. 

\subsection{Signed Boundary Decomposition}

A signed decomposition of a cell $c$ into simplices $\pm\Delta_1,\dotsc,\mp\Delta_n$
is a linear combination of simplices $\Delta_1,\ldots,\Delta_n$ such that
\begin{equation}
   c = \alpha_1\Delta_1 + \cdots + \alpha_n\Delta_n
\end{equation}
where $\alpha_i= {+1}$ if the simplex is added and $\alpha_i={-1}$ if it is subtracted. 
As the order in which we add or subtract simplices does not matter 
one might also call this ``Commutative Constructive Solid Geometry'' (CCSG).

The algorithms presented here operate on the relational representation of a complex 
but we now take its ``classical'' view where each $C_i$ is the free Abelian group of cells in $X_i$, 
and the boundary operators $\partial_i$ are linear maps between them satisfying 
$\partial_{i-1}(\partial_{i}(x)) = 0$ for all $x\in C_i$.
Now assume such complex
\begin{equation}
\xymatrix{
 \mathcal{C} : C_n     \ar[r]^{\partial_{n}} 
             & C_{n-1} \ar[r]^{\partial_{n-1}}
             & \cdots  \ar[r]^{\partial_{1}}
             & C_0}
\end{equation}
be given. 

To triangulate  an $n$-cell $c_n$ from $C_n$ (hence $c_n\in X_n$) we simply add an arbitrary point $a$ 
from the interior of $c_n$ as new vertex $a$ (``apex'') to the vertices and replace our cell 
$c_n$ by a ``cone'' $a\otimes\partial(c_n)$ over its boundary with apex $a$. 
The tensor operator $a\otimes x$ is simply the concatenation of $a$ and $x$, either by creating a pair 
$(a,x)$ or by appending tuples:
\begin{align}
\nonumber
a\otimes\partial(c_n) &= a\otimes(\alpha_1\cdot d_1 + \cdots + \alpha_m\cdot d_m) \\
                      &= a\otimes\alpha_1\cdot d_1 + \cdots + a\otimes\alpha_m\cdot d_m\\
\nonumber
                      &= \alpha_1\cdot a\otimes d_1 + \cdots + \alpha_m\cdot a\otimes d_m
\end{align}
The boundary of one such such element $a\otimes d_i$ can be defined as 
\begin{align}\label{eqn:cone-boundary}
\partial(a\otimes d_i) := d_i - a\otimes\partial(d_i).
\end{align}
This is a boundary operator, indeed. It slightly modifies the 
Eilenberg-Zilber-formula \cite{EilenbergZilber:Product} for tensor product boundaries
\begin{align}
\delta(a\otimes x) := \delta(a)\otimes x + (-1)^{\dim a} a \otimes\delta(x)
\end{align}
to 
\begin{align}
\partial(a\otimes x) := \partial(a)\otimes x - (-1)^{\dim a} a \otimes\partial(x)
\end{align}
and specifies $\partial(a)=\langle\rangle$ which means that the boundary of the apex $a$ is the empty 
tuple---the identity element of concatenation: $\langle\rangle\otimes x = x = x \otimes \langle\rangle$. 
Subtraction instead of addition is achieved by simply ``shifting'' the dimension of $a$ from zero to 1 and 
considering $\langle\rangle$ having dimension 0. 

In contrast to the original, our variant of the Eilenberg-Zilber-formula is compatible with the 
simplicial boundary operator which we denote here by $\delta$:
\begin{align}
\nonumber
\partial&(a \otimes \simplex{b,c,d,\ldots}) \\
\nonumber
        & =  \simplex{b,c,d,\ldots} - \simplex{a,c,d,\ldots} + \simplex{a,b,d,\ldots}-\simplex{a,b,c,\ldots}+\cdots\\
\label{eqn:simplicial-eilenberg-zilber}
        &=  \delta(\simplex{a,b,c,d,\ldots})
\end{align}
This even works for arbitrary $n$\nobreakdash-simplices $\simplex{a_0,\dotsc,a_n}\otimes\simplex{b,c,d,\ldots}$.

We remind that the simplicial boundary $\delta$ of a simplex $\simplex{v_0,\dotsc,v_n}$ is defined as 
\begin{equation}
  \delta(\simplex{v_0,\dotsc,v_n}) := \sum_{i=0}^n (-1)^{i}\simplex{v_0,\dotsc,v_{i-1},\ v_{i+1},\dotsc,v_n}.
\end{equation}

Our ``cone'' over the boundary in fact has the same algebraic boundary as the original cell:
\begin{align}
\nonumber
\partial(a \otimes \partial(c_n)) 
        &=\partial(c_n) - a\otimes\partial(\partial(c_n)) =\partial(c_n) - a\otimes 0\\
        &=\partial(c_n).
\end{align}
So a property of $c_n$ computed via its boundary can also be computed by the proposed 
triangulation---even when $a$ is outside $c_n$ or a vertex of the boundary of $c_n$.

\begin{example}[Triangulating a Square]
The left hand side below shows a square $S$ in counter-clockwise orientation (as indicated by the bent 
arrow $\xymatrix@1{\ar@{.>}[r]&}$ and the right hand side its triangulation with a newly introduced apex ``$0$'':
$$
\xymatrix{\\
  \bullet
  \ar[rr]^(-.05)1^(1.05)2^{12}_{\displaystyle -}
  \ar[dd]_{13}^{\displaystyle +}
                 &   &    \bullet 
                          \ar[dd]^{24}_{\displaystyle -}
                                       & & &   \bullet
                                              \ar[rr]^(-.05)1^(1.05)2^{12}|{}="e12"
                                              \ar[dd]_{13}|{}="e13"
                                                           &   & \bullet \ar[dd]^{24}|{}="e24"\\
                 & S 
                     \ar@{}[dl]|(0.25){}="DL"\ar@{}[dr]|(0.25){}="DR"\ar@{}[ur]|(0.25){}="UR"
                     \ar@{.}"DL";"DR"\ar@{.>}"DR";"UR"
                     &                 &\ar[r]^{\displaystyle\mu} 
                                         & &                & \,\bullet_0 \ar@{}[]|{}="apex"
                                                               \ar[lu]|(.55)*+={\scriptstyle 01}
                                                               \ar[ru]|(.55)*+={\scriptstyle 02}
                                                               \ar[ld]|(.55)*+={\scriptstyle 03}
                                                               \ar[rd]|(.55)*+={\scriptstyle 04}
                                                               \ar@{}[];"e12"|(.7)*+<2pt>[F.]{\scriptstyle -012}
                                                               \ar@{}[];"e13"|(.6)*+<2pt>[F.]{\scriptstyle +013}
                                                               \ar@{}[];"e24"|(.6)*+<2pt>[F.]{\scriptstyle -024}
                                                               &     \\
  \bullet  \ar[rr]_(-.05)3_(1.05)4_{34}^{+}
                 &   &    \bullet          & & &   \bullet \ar[rr]_(-.05)3_(1.05)4_{34}|{}="e34"
                                              \ar@{}"apex";"e34"|(.7)*+<2pt>[F.]{\scriptstyle +034}
                                                               &   & \bullet}
$$
The signs of the four triangles correspond to their orientation: If a triangle has clock-wise orientation 
(like $\simplex{0,1,2}$ which visits its vertices $0$, $1$, and $2$ in clock-wise order) 
it gets a negative sign. 
Each common edge of two adjacent signed triangles vanish when the boundary of their sum is computed. 
In 
$\partial(\simplex{0,1,3}-\simplex{0,1,2})=+\simplex{1,3}-\simplex{0,3}+\simplex{0,2}-\simplex{1,2}+0\times\simplex{0,1}$, 
for example, the common edge $\simplex{0,1}$ cancels out. 

The boundary of the left hand side complex is
\begin{align}
  \partial_2(S) = -\simplex{1,2} + \simplex{1,3} - \simplex{2,4} + \simplex{3,4}
\end{align}
for the face and 
\begin{align}
\nonumber
  \partial_1(\simplex{1,2}) &= \simplex{2}-\simplex{1} &
  \partial_1(\simplex{1,3}) &= \simplex{3}-\simplex{1}\\
  \partial_1(\simplex{2,4}) &= \simplex{4}-\simplex{2} &
  \partial_1(\simplex{3,4}) &= \simplex{4}-\simplex{3}
\end{align}
for the edges. The triangulation $\mu(S)$ then is:
\begin{align}
\nonumber
  \mu(S) = "0"\otimes\partial_2(S) 
              &= "0"\otimes(-\simplex{1,2} + \simplex{1,3} - \simplex{2,4} + \simplex{3,4})\\
              &=-\simplex{0,1,2} + \simplex{0,1,3} - \simplex{0,2,4} + \simplex{0,3,4}
\end{align}
Now, algebraically, $\mu(S)$ has the same boundary as $S$:
\begin{align}
\nonumber
\partial(\mu(S)) &= \partial(-\simplex{0,1,2} + \simplex{0,1,3} - \simplex{0,2,4} + \simplex{0,3,4})\\
\nonumber
               &= -\partial\simplex{0,1,2} + \partial\simplex{0,1,3} - \partial\simplex{0,2,4} + \partial\simplex{0,3,4})\\
\nonumber
               &= -(\simplex{1,2}-\simplex{0,2}+\simplex{0,1}) + (\simplex{1,3}-\simplex{0,3}+\simplex{0,1}) \\
\nonumber
               &\phantom{=\ } -(\simplex{2,4}-\simplex{0,4}+\simplex{0,2}) + (\simplex{3,4}-\simplex{0,4}+\simplex{0,3})\\
\nonumber
               & =-\simplex{1,2} + \simplex{1,3} - \simplex{2,4} + \simplex{3,4}\\
               &=\partial_2(S).
\end{align}
Each summand $\pm\simplex{0,x}$ in the above sum where the simplex starts with $"0"$ has a 
simplex $\mp\simplex{0,x}$ such that both cancel to zero. 
\end{example}

This approach, however, has a shortcoming: introducing a new vertex 
for each cell means that each sequence $(c_n,c_{n-1},\dotsc,c_0)$ of incident cells---a so-called 
\emph{cell tuple}---is represented by one simplex. 
But this can result in extremely many simplices---the number of cell tuples then grows more than 
exponentially with the dimension of the complex. 
This is also true for our approach proposed next, but at least it greatly reduces the number of simplices. 

\subsubsection{Modifying Cohen and Hickey's Approach}

To ease the problem of the big number of simplices we use an existing boundary vertex instead of 
providing a new vertex for each cell. 
Then all boundary cells incident with that vertex degenerate, have a zero boundary, and can be dropped. 
We will now iteratively construct a signed simplicial decomposition as a morphism $\mu = (\mu_n,\dotsc,\mu_0)$ 
from our polytope complex $\mathcal{C}$ to a simplicial complex\footnote{with some salt added: As the simplices may overlap 
it can be disputed if this really qualifies as a ``simplicial complex''. Anyhow, it is a complex of simplices with the 
simplicial boundary operator.} dimension by dimension starting with dimension zero, 
This morphism is then the desired signed simplicial decomposition. 

We still denote the $i$-cells of our complex by $X_i$. 
At the first step the vertices $X_0$ in $C_0$ are simply labelled with integer numbers by $\mu_0$ which gives a bijection 
\begin{equation}
  \mu_0 : X_0 \to [1,\dotsc, \#X_0]\subseteq\bbbn
\end{equation}
and its linear continuation to $C_0\to\bbbz[1,\dotsc,\#X_0]$. 
This specifies a linear order $<$ on the vertices and thereby a priori fixes an 
orientation of every simplex. 
The order is defined as $\ a < b :\Longleftrightarrow \mu_0(a) < \mu_0(b)\ $ for all $a,b\in X_0$. 
We leave it open how permuting the labels can affect the size of the decomposition result and how we then can 
find an optimal $\mu_0$. 

Each $i$\nobreakdash-simplex is then an ascending sequence of $i+1$ different vertices.
Our assumptions on the polytope imply that each edge $e\in X_1$ has two different boundary vertices $a$ 
and $b$ and one of them is the starting vertex and the other is the ending vertex. This gives 
two matrix entries $D_1(e,a) = -D_1(e,b)$ in the boundary matrix $D_1$. 
An entry $+1$ indicates an ending vertex and $-1$ stands for the starting vertex. 
This matrix entry of the maximal vertex will become the new sign of our image edge going from minimal to maximal 
vertex. Hence $\mu_1$ is defined by 
\begin{equation}
  \mu_1 : X_1 \to \bbbz[0,\dotsc, \#X_0]^2, e \mapsto \begin{cases}
                                                           D_1(e,b)\times\simplex{\mu_0(a), \mu_0(b)} &: a < b\\
                                                           D_1(e,a)\times\simplex{\mu_0(b), \mu_0(a)} &: b < a\;.
                                                           \end{cases}
\end{equation}
This function inverts the edge orientation and the sign when its vertices are ordered against their total ordering 
imposed by $\mu_0$ and guarantees that the first vertex in the simplex $\simplex{a,b}$ is always minimal. 
Note that the first two vertices are always different. 
Now this guarantee will be kept throughout every dimension. Here we denote the $i^{\text{th}}$ simplicial boundary 
by $\delta_i$.

We now show that our partially finished signed decomposition $(\mu_1,\mu_0)$ is a morphism from the $1$-skeleton 
$(X_1,X_0)$ to a simplicial complex $\Delta$ by showing  
\begin{equation}\label{eqn:mu-o-partial-commutes}
   \delta_1\circ\mu_1 = \mu_0\circ\partial_1\;.
 \end{equation}

Let $e$ be an edge in $X_1$, $a$ its starting vertex, and $b$ be the ending vertex. This means 
$\partial_1(e)=b-a$. and  $D_1(e,b)={+1}$ and $D_1(e,a)={-1}$. Then for the right hand side of 
Equation~\eqref{eqn:mu-o-partial-commutes} we have:
\begin{align}\label{eqn:muopartial}
   \mu_0(\partial_1(e)) &=\mu_0(b-a) =\mu_0(b) -\mu_0(a)
\intertext{In case $a<b$ we have for the left-hand side:}
\nonumber
     \delta_1(\mu_1(e))&=\delta_1(D_1(e,b)\times\simplex{\mu_0(a),\mu_0(b)})=\delta_1\simplex{\mu_0(a),\mu_0(b)}\\
                       &=\mu_0(b) -\mu_0(a)\;.
\intertext{In case $b<a$ the left hand side gives:}
\nonumber
     \delta_1(\mu_1(e))&=\delta_1(D_1(e,a)\times\simplex{\mu_0(b),\mu_0(a)})=\delta_1(-\simplex{\mu_0(b),\mu_0(a)})\\
\nonumber              &=-\delta_1(\simplex{\mu_0(b),\mu_0(a)}) = -(\mu_0(a) -\mu_0(b))\\
                       &=\mu_0(b) -\mu_0(a)\;.
\end{align}
In both cases we have $\mu_0(\partial_1(e))=\delta_1(\mu_1(e))$ for every edge $e$ and therefore 
Equation~\eqref{eqn:mu-o-partial-commutes} holds. 

The higher dimensional maps $\mu_2,\mu_3,\ldots$ can now be computed iteratively each of them using its previously 
accomplished predecessor: 
Let $\mu_i$ be given. Then we use it to compute $\mu_{i+1}$ from $\mu_i$ and $\partial_{i+1}$ for each 
$i+1$\nobreakdash-cell $c$ in $C_{i+1}$:
First take the given boundary $\partial_{i+1}(c)$ which is a linear combination 
\begin{equation}
\partial_{i+1}(c) = \alpha_1 d_1 + \dotsb + \alpha_k d_k
\end{equation}
where the $\alpha_j$ are the entries $D_{i+1}(c,d_j)$ in the boundary matrix $D_{i+1}$ of $\partial_{i+1}$.
Then compute the image of $\partial_{i+1}(c)$ under the morphism $\mu_i$ which gives 
\begin{equation}\label{eqn:mu-d-prev}
\mu_i\circ\partial_{i+1}(c) = \alpha_1 \mu_i(d_1) + \dotsb + \alpha_k \mu_i(d_k).
\end{equation}
We know that each simplex $\mu_i(d_j)$ in above Equation~\eqref{eqn:mu-d-prev} is a sequence $\simplex{\nu_0,\dotsc,\nu_i}$ 
of vertex numbers (given by $\mu_0$) in strictly ascending order. Therefore the first vertex number $\nu_0$ is minimal in that 
simplex. We now take the minimal number $a$ of all these vertices in the boundary of $c$ and define 
\begin{align}\label{eqn:a-tensor-mu-d-prev}
   a\otimes\mu_i\circ\partial_{i+1}(c) = \alpha_1\times a\otimes\mu_i(d_1) + \dotsb + \alpha_k\times a\otimes\mu_i(d_k)
\end{align}
which is a linear combination of simplices $\simplex{a,\nu_0,\dotsc,\nu_i}$ which are the $i$-simplices of the triangulated 
boundary of $c$ with the minimal vertex $a$ attached at front to get a new $i+1$ simplex. 

The operator $a\otimes\mu_i\circ\partial_{i+1}$, in fact, is a morphism:
\begin{align} 
\nonumber
  \delta_{i+1}&(a\otimes\mu_i\circ\partial_{i+1}(c))\\
\nonumber
       &= \mu_i\circ\partial_{i+1}(c) - a\otimes\delta_{i}(\mu_i\circ\partial_{i+1}(c)) 
                     &&\text{\quad by Eqn.~\eqref{eqn:cone-boundary}}\\
\nonumber
       &= \mu_i\circ\partial_{i+1}(c) - a\otimes\delta_{i}\circ\mu_i\circ\partial_{i+1}(c)\\
\nonumber
       &= \mu_i\circ\partial_{i+1}(c) - a\otimes\mu_{i-1}(\partial_{i}\circ\partial_{i+1}(c))
                     &&\text{\quad by $\mu_i$ being a morphism}\\
\nonumber
       &= \mu_i\circ\partial_{i+1}(c) - a\otimes\mu_{i-1}\circ\hat{0}(c)
                     &&\text{\quad by $\partial$ being a boundary}\\
\nonumber
       &= \mu_i\circ\partial_{i+1}(c) - a\otimes 0\\
       &= \mu_i\circ\partial_{i+1}(c).
\end{align}
Now each boundary simplex $\simplex{\nu_1,\dotsc,\nu_n}$ which contains $a$ has that vertex $a$ at its front because 
it is the minimal vertex of all boundary simplices and $\nu_1$ is the minimal vertex of that simplex. 
Hence $a=\nu_1$ and thus the simplex degenerates to $\simplex{a,a,\nu_2,\dotsc,\nu_n}$. 
But then every non-zero summand of its simplicial boundary also starts with $\simplex{a,a,\dotsc}$. 

But such an element does not occur in $\delta_i\circ\mu_{i-1}(c)$ because all vertices in all simplices in $\mu_{i-1}(c)$ 
are different by precondition. By the morphism property $\delta_i\circ\mu_{i-1}(c) = \mu_i\circ\partial_{i+1}(c)$ these 
elements can occur in $\mu_i\circ\partial_{i+1}(c)$ only with a zero coefficient 
and therefore can be removed from $\mu_i\circ\partial_{i+1}(c)$ without (algebraically) modifying the boundary 
$\delta_{i+1}$. 

Our new morphism $\mu_{i+1}$ does exactly this and thus guarantees the precondition that all vertices 
of a simplex are different:
\begin{equation}
\mu_{i+1}(c):=a_c\otimes\mu^{(a_c)}_i\circ\partial_{i+1}(c)
\end{equation}
where $a_c$ is the minimal 
vertex touching $c$ and $\mu^{(a_c)}_i$ is $\mu_i$ after having removed all simplices containing $a_c$.

\begin{example}[Triangulating our Square Again]
The left hand side diagram shows a square $S$ and the right hand side its triangulation.
We do this step by step starting with $\mu_0$ which enumerates the vertices in an arbitrary manner:

\ \xymatrix{\\
  \bullet
  \ar[rr]^(-.05)a^(1.05)b^{ab}_{\displaystyle -}
  \ar[dd]_{ac}^{\displaystyle +}
                 &   &    \bullet 
                          \ar[dd]^{bd}_{\displaystyle -}
                                       & & &   \bullet
                                              \ar@{}[rr]^(-.05)4^(1.05)2
                                                           &   & \bullet\\
                 & S &                 &\ar[r]^{\displaystyle\mu_0} 
                                         & &                &  &     \\
  \bullet  \ar[rr]_(-.05)c_(1.05)d_{cd}^{+}
                 &   &    \bullet      & & &   \bullet 
                                               \ar@{}[rr]_(-.05)1_(1.05)3
                                                               &   & \bullet \\
&}

Now we can compute the edges which run from lower to higher vertex number:

 \ \xymatrix{\\
   \bullet
   \ar[rr]^(-.05)a^(1.05)b^{ab}_{\displaystyle -}
   \ar[dd]_{ac}^{\displaystyle +}
                  &   &    \bullet 
                           \ar[dd]^{bd}_{\displaystyle -}
                                        & & &   \bullet
                                               \ar@{<-}[rr]^(-.05)4^(1.05)2^{-\simplex{2,4}}|{}="e24"
                                               \ar@{<-}[dd]_{-\simplex{1,4}}|{}="e14"
                                                            &   & \bullet \ar[dd]^{+\simplex{2,3}}|{}="e23"\\
                  & S &                 &\ar[r]^{\displaystyle\mu_1} 
                                          & &                &  &     \\
   \bullet  \ar[rr]_(-.05)c_(1.05)d_{cd}^{+}
                  &   &    \bullet          & & &   \bullet 
                                                    \ar[rr]_(-.05)1_(1.05)3_{+\simplex{1,3}}|{}="e34"
                                                                &   & \bullet \\
 &}
 
Remember, the the boundary of $S$ is $\partial_2(S) = cd - bd - ab + ac$. Then 
\begin{align}
\nonumber
\mu_1(\partial_2(S)) &= \mu_1(cd) - \mu_1(bd) - \mu_1(ab) + \mu_1(ac)\\
\nonumber
                   &= \simplex{1,3} - \simplex{2,3} - (-\simplex{2,4}) + (-\simplex{1,4})\\
                   &= \simplex{1,3} - \simplex{2,3} + \simplex{2,4} -\simplex{1,4}
\intertext{
We take the minimal vertex $1$, remove all simplices that start with $1$, and then append $1$ to 
the remaining simplices:}
\mu_2(S) &= -\simplex{1,2,3} + \simplex{1,2,4}
\end{align}

 \ \xymatrix{\\
   \bullet
   \ar[rr]^(-.05)a^(1.05)b^{ab}_{\displaystyle -}
   \ar[dd]_{ac}^{\displaystyle +}
                  &   &    \bullet 
                           \ar[dd]^{bd}_{\displaystyle -}
                                        & & &   \bullet
                                               \ar@{<-}[rr]^(-.05)4^(1.05)2^{\simplex{2,4}}|(0.2){}="e24"
                                               \ar@{}"e24";[dr]|(.4)*+<2pt>[F.]{\scriptstyle+\simplex{1,2,4}}
                                               \ar@{<-}[dd]_{\simplex{1,4}}
                                                             &  & \bullet \ar[dd]^{\simplex{2,3}}\\
                  & S &                 &\ar[r]^{\displaystyle\mu_2} 
                                          & &                & 
                                                                &     \\
   \bullet  \ar[rr]_(-.05)c_(1.05)d_{cd}^{+}
                  &   &    \bullet          & & &   \bullet 
                                                    \ar[rr]_(-.05)1_(1.05)3_{\simplex{1,3}}|(.8){}="e13"
                                                    \ar@{}"e13";[ur]|(.4)*+<2pt>[F.]{\scriptstyle-\simplex{1,2,3}}
                                                    \ar@{-->}[uurr]|{\simplex{1,2}}
                                                             &   & \bullet \\
 &}\\
Note that the $2$\nobreakdash-simplex (or triangle) $\simplex{1,2,4}$ is oriented counter-clockwise just as the face 
$S$ and therefore gets a positive sign whereas triangle $\simplex{1,2,3}$ is oriented clockwise and accordingly gets a 
negative sign to indicate that is must be flipped to be compatible with the orientation of $S$.
\end{example}

Finally the simplices, which are still tuples $\simplex{\nu_0,\dotsc,\nu_i}$ of natural numbers are considered 
simplices in $\bbbr^n$ by application of $\mu_0^{-1}$, the inverse of $\mu_0$. This gives 
$\simplex{v_0,\dotsc,v_i} = \simplex{\mu_0^{-1}(\nu_0),\dotsc,\mu_0^{-1}(\nu_i)}$.

Now the $n$-d ``commutative CSG'' is defined as follows: If the sign of the $n$-simplex orientation matches the sign 
assigned to it by the morphism the (absolute) simplex volume will be added to the polytope cell whereas in case that 
orientation sign and assigned sign are different the (absolute) simplex volume will be removed from the polytope. 
``Simplex orientation'' of an $n$-simplex in $\bbbr^n$ can easily be computed: 
The determinant 
\begin{equation}
\det(\simplex{v_0,\dotsc,v_n}) =
  \left|\ \begin{array}{c@{\quad}c@{\quad}c@{\quad}c}
      x_1-x_0 & x_2-x_0 & \cdots & x_n-x_0 \\
      y_1-y_0 & y_2-y_0 & \cdots & y_n-y_0 \\
      \vdots  & \vdots  & \ddots & \vdots \\
      z_1-z_0 & z_2-z_0 & \cdots & z_n-z_0
    \end{array}
  \ \right|
\end{equation}
has $n!$ times the signed volume of the corresponding simplex and the sign is the orientation. 
The values $x_i,y_i,\ldots,z_i$ are the $n$ vertex coordinates of vertex $v_i$. 

\section{Simplex Intersection}

One application of such signed decomposition is polytope intersection or---more generally---polytope complex 
intersection: the pair-wise intersection of cells in a polytope complex which itself gives a new polytope complex. 

Wit those signed simplices $\mu$ we compute the polytope complex intersection of, say, 
$\mathcal{C}$ and $\mathcal{K}$. But how do we intersect simplices?

\subsection{Application of the Active-Set-Method}

Assume there are two simplices: an $n$\nobreakdash-simplex $a = \simplex{a_0,\ldots,a_n}$ and an 
$m$\nobreakdash-simplex $b = \simplex{b_0,\ldots,b_m}$. 
We will give a brief sketch of how the cells of the intersection complex of $a\cap b$ can be found using a 
modified version of the active-set-method \cite{Nocedal:Optimization} a well-known numerical optimisation technique. 

Each point $p$ in a simplex $\simplex{a_0,\dotsc,a_n}$ has convex coordinates $(\alpha_0,$ $\dotsc,\alpha_n)$ 
such that $p = \alpha_0 a_0 + \dotsb + \alpha_n a_n$. In other words, these coordinates must sum up to 
$1$ and neither is negative. Let $q = \beta_0 b_0 + \dotsb + \beta_m b_m$ be such point in the other 
simplex $b$.
Then the pair of points $(p,q)$ can be expressed by concatenating 
their barycentric coordinates to a vector $\vec{x} = (\alpha_0,\dotsc\alpha_n,\,\beta_0,\dotsc,\beta_m)$. 
It represents an intersection iff (the square of) the distance of $p$ and $q$ is zero. 
The square of the distance can be computed by inserting $\vec{x}$ into 
\begin{equation}
d^2(\vec{x}) = 
  \vec{x}\cdot\left(\begin{array}{cccccccccccccccccc}
    a_0a_0 & \cdots & a_0a_n && -a_0b_0 & \cdots & -a_0b_m \\
    \vdots & \ddots & \vdots && \vdots & \ddots & \vdots  \\
    a_na_0 & \cdots & a_na_n && -a_nb_0 & \cdots & -a_nb_m \\[\jot]
    -b_0a_0 & \cdots & -b_0a_n && b_0b_0 & \cdots & b_0b_m \\
    \vdots &  \ddots & \vdots && \vdots & \ddots & \vdots  \\
    -b_ma_0 & \cdots & -b_ma_n && b_mb_0 & \cdots & b_mb_m 
  \end{array}\right) \cdot \vec{x}^T
\end{equation}
a so-called ``quadratic form''. The matrix entries are the dot product of the vertices in $\bbbr^n$.
To minimise this equation, subject to the constraints 
\begin{equation}\label{eqn:affine}
\alpha_0 + \dotsb + \alpha_n\ = {\quad 1\quad} = \ \beta_0 + \dotsb + \beta_m ,
\end{equation}
the following Karush-Kuhn-Tucker-system (or KKT-system) must be solved:
\begin{equation}\label{eqn:kkt-system}
\left(\begin{array}{cccccccccccccccccc}
     a_0a_0 & \cdots & a_0 a_n   & -a_0b_0 & \cdots & -a_0b_m & 1 && 0\\
     \vdots & \ddots & \vdots   & \vdots & \ddots & \vdots & \vdots&& \vdots  \\
     a_na_0 & \cdots & a_na_n   & -a_nb_0 & \cdots & -a_nb_m & 1 && 0\\[\jot]
    -b_0a_0 & \cdots & -b_0 a_n & b_0b_0 & \cdots & b_0b_m & 0 && 1\\
    \vdots  &  \ddots & \vdots  & \vdots & \ddots & \vdots & \vdots&& \vdots \\
    -b_ma_0 & \cdots & -b_m a_n & b_mb_0 & \cdots & b_mb_m & 0 && 1\\[\jot]
         1  & \cdots &    1     &  0    & \cdots &     0   & 0 && 0\\
         0  & \cdots &    0     &  1    & \cdots &     1   & 0 && 0\\
  \end{array}\right) 
\cdot
\left(\begin{array}{cccccccccccccccccc}
     \alpha_0\\
     \vdots \\
     \alpha_n\\[\jot]
     \beta_0\\
    \vdots\\
    \beta_m\\[\jot]
    \Lambda_a\\
    \Lambda_b\\
  \end{array}\right)
=
\left(\begin{array}{cccccccccccccccccc}
     0\\
     \vdots \\
     0\\[\jot]
     0\\
    \vdots\\
    0\\[\jot]
    1\\
    1\\
  \end{array}\right)
\end{equation}

Note that the last two lines of this equation system are Equations~\eqref{eqn:affine}.

Now a vertex is called \emph{active} at $\vec{x}$ if its corresponding barycentric coordinate is zero, 
and otherwise \emph{inactive}. 
A vector is \emph{feasible} iff it represents a convex combination of the vertices, hence iff 
none of the barycentric coordinates is negative. 
A set of vertices is \emph{feasible} if its KKT-system has a unique feasible solution. 

\begin{samepage}
To find all intersection cells of the two simplices we first compute the intersection vertices with the following 
algorithm:
\begin{itemize}
\item Initialise a set $S = \mathpset{\mathset{a_i,b_j}}{i = 0\dotsc n, j = 0\dotsc m}$, the initially inactive 
      sets, and initialise an empty set $R$ (the ``Result''). 
\item While $S$ is not empty do:
  \begin{itemize}
    \item Chose one inactive set $I$ from $S$ and remove it from $S$. 
    \item Restrict the KKT-system of Equation~\eqref{eqn:kkt-system}  to $I$ and solve it, say, by calling 
          LAPACK's \texttt{dgesv}-routine \cite{LAPACK:UGuide}. 
          ``Restrict'' means  ``remove each row and column $x$ which is not in $I$ from the equation system'' 
          (or, equivalently, set each such row and column $x$ in the matrix to $0$ and then the diagonal 
          entry at $x$ to $1$).
    \item If the solution is unfeasible the intersection point is outside of one of the simplices. 
          Then discard $I$ and continue the loop with the next $I$ from $S$. 
    \item Otherwise, if the Lagrangian multipliers $\Lambda_a$ and $\Lambda_b$ of the solution are both zero,
          the corresponding simplices intersect at the point with these barycentric coordinates. 
          Then add $I$ to $R$, because $I$ then indicates an intersection vertex of the simplex intersection. 
    \item If the solution is feasible but at least one of the Lagrangians is not zero the simplices are either 
          skew or parallel and their affine spaces do not intersect. 
          Then multiply the solution with the Matrix of the unrestricted KKT-system. 
          This gives (half) the gradient of the (squared) distance function at $\vec{x}$. 
          Then for each vertex $v_i\notin I$ where that gradient is negative add $I+\mathset{v_i}$ to $S$. 
    \end{itemize}
  \item When $S$ is empty return result $R$.
\end{itemize}
\end{samepage}

We can prove that this algorithm always terminates and computes every mini\-mal inactive set for each vertex of 
the simplex intersection. We will not carry out that proof here. It consists in showing that every such inactive 
set has a sequence of vertices which can be removed one by one without rendering the remaining set unfeasible until 
it becomes one of the initially inactive sets. 

A forthcoming paper will show how we handle ill-conditioned matrices and ``critical'' intersections like 
an intersection of a vertex with an edge of a triangle. 

Each set $I$ in the above computed set $R$ is a set of vertices denoting a pair of sides one from each simplex. 
This pair of sides intersects in one unique point. To get the higher-dimensional intersection cells we 
simply compute all possible unions of these inactive sets. 

The following diagram shows an intersection of 
two triangles on the left hand side together with the inactive sets of the intersection vertices on the 
right-hand side: 
\begin{center}
\begin{picture}(320,120)(0,0)
\put( 40, 20){\circle*{4}}   \put( 35, 10){$a_1$}
\put(120, 20){\circle*{4}}   \put(115, 10){$a_2$}
\put( 80,100){\circle*{4}}   \put( 75,110){$a_3$}
\put( 40, 20){\line(1,0){80}}
\put( 40, 20){\line(1,2){40}}
\put(120, 20){\line(-1,2){40}}

\put( 20,40){\circle*{4}}    \put( 15,30){$b_4$}
\put(140,40){\circle*{4}}    \put(135,30){$b_5$}
\put( 80,80){\circle*{4}}    \put( 75,85){$b_6$} 
\put( 20,40){\line(1,0){120}}
\put( 20,40){\line(3,2){60}}
\put(140,40){\line(-3,2){60}}

\put(200, 20){\circle*{2}}   \put(195, 12){\scriptsize{1}}
\put(280, 20){\circle*{2}}   \put(281, 12){\scriptsize{2}}
\put(240,100){\circle*{2}}   \put(238,105){\scriptsize{3}}
\put(200, 20){\line(1,0){80}}
\put(200, 20){\line(1,2){40}}
\put(280, 20){\line(-1,2){40}}

\put(180,40){\circle*{2}}    \put(188, 41){\scriptsize{4}}
\put(300,40){\circle*{2}}    \put(288, 41){\scriptsize{5}}
\put(240,80){\circle*{4}}    \put(238, 70){\scriptsize{6}}

\put(180,40){\line(1,0){120}}
\put(180,40){\line(3,2){60}}
\put(300,40){\line(-3,2){60}}

\put(240,80){\circle*{4}}   \put(240,83){\scriptsize{1236}}
\put(225,70){\circle*{4}}   \put(206,70){\scriptsize{1346}}
\put(255,70){\circle*{4}}   \put(258,70){\scriptsize{2356}}

\put(210,40){\circle*{4}}   \put(190,33){\scriptsize{1345}}
\put(270,40){\circle*{4}}   \put(273,33){\scriptsize{2345}}
\end{picture}
\end{center}
The vertex $\mathset{1,3,4,6}$ indicates that it is on the intersection of edge $\simplex{1,3}$ 
with edge $\simplex{4,6}$, whereas $\mathset{1,2,3,6}$ denotes the intersection of vertex $6$ with 
triangle $\simplex{1,2,3}$. 
Note that $\mathset{1,3,4,5}\cup\mathset{2,3,4,5}=\mathset{1,2,3,4,5}$---the union of the inactive sets of the 
two bottom vertices---gives the inactive set of the bottom line between these vertices. If we restrict the KKT-system 
to that set then its solution space is of dimension one. It contains the barycentric coordinates 
of all points in the affine space of that edge, whereas the feasible solutions denote the points on that edge itself.

\subsection{The Algebraic Simplex Intersection Boundary}

From now on we will write the sets $\mathset{1,2,3,5}$ as $"1235"$, i.e.\ a string 
of integers. 
What follows is the set of all unions of $"1345"$, $"2345"$, $"1346"$, $"2356"$, and $"1236"$:
\begin{align}
\nonumber
  X = \{&"1345", "2345", "1346", "2356", "1236",       &&\text{the vertices}\\
        &"12345", "13456", "23456", "12346", "12356",  &&\text{the edges}\\
\nonumber
        &"123456"\}                                    &&\text{the face}
\end{align}
This gives a topological data type $(X,\supseteq)$ where each set is bounded by its subsets in $X$. 
To get a relational complex we need an algebraic  boundary operator. A possible solution could be to 
restrict the simplicial boundary to $X$. This gives indeed a complex boundary 
(see also \cite{Lefschetz:ComplexIntersection}) but none which is useful for our purpose. 
We will indicate restriction to $X$ by either ${<}\text{expr}{>}|_X$ or by striking 
\sout{out} sets which are not in $X$. 
Let us try that simple approach: 
Computing some edge boundary gives strange edges like
\begin{align}
\nonumber
\partial|_X("23456") &=  + \text{"\sout{3456}"} - \text{"\sout{2456}"}  +  "2356" - \text{"\sout{2346}"} + "2345" \\
                     &= + "2356" + "2345".
\end{align}
Remember that a positive sign indicates ``end point'', and a negative sign is attached to a starting point. 
So we see that, instead of assigning a starting point and an endpoint, 
that boundary operator assigns \emph{two} endpoints ``2356'' and ``2345'' and \emph{no} starting point to the 
edge ``23456''. 
Hence, that boundary defines no orientation. 
Computing the boundary of the face and all edges gives the following result:

\begin{center}
\setlength{\unitlength}{1.4\unitlength}
\begin{picture}(220,100)(0,0)


\put(20, 20){\circle*{1}}   
\put(100, 20){\circle*{1}}  
\put(60,100){\circle*{1}}   

\put(30, 20){\line( 1,0){60}}  
\put(26, 32){\line( 1,2){24}}  
\put(94,32){\line(-1,2){24}}  

\put(0,40){\circle*{1}}    
\put(120,40){\circle*{1}}   

\put(20,40){\line(1,0){80}}  
\put(18,52){\line(3,2){42}}   
\put(102,52){\line(-3,2){42}} 

\put(60,80){\circle*{3}}   \put(55,83){\scriptsize{1236}}
\put(45,70){\circle*{3}}   \put(29,70){\scriptsize{1346}}
\put(75,70){\circle*{3}}   \put(78,70){\scriptsize{2356}}

\put(30,40){\circle*{3}}   \put(16,35){\scriptsize{1345}}
\put(90,40){\circle*{3}}   \put(94,35){\scriptsize{2345}}

\put(75,70){\vector(1,-2){14}} \put(90,40){\vector(-1,2){14}} 
\put(60,80){\vector(3,-2){13}} 
\put(60,80){\vector(-3,-2){6}} \put(45,70){\vector(3,2){6}} 
\put(45,70){\vector(-1,-2){14}} 
\put(30,40){\vector(1,0){58}}   

\put(50,53){\scriptsize{123456}}
\put(54,35){\scriptsize{12345}}  \put(58,43){\scriptsize{--}}
\put(85,53){\scriptsize{23456}} \put(76,53){\scriptsize{+}}
\put(21,53){\scriptsize{13456}} \put(40,53){\scriptsize{--}}

\put(69,78){\scriptsize{12356}} \put(63,71){\scriptsize{--}}
\put(39,78){\scriptsize{12346}} \put(51,71){\scriptsize{+}}


\put(180,80){\circle*{2}}  
\put(165,70){\circle*{2}}  
\put(195,70){\circle*{2}}  
\put(150,40){\circle*{2}}  
\put(210,40){\circle*{2}} 

\put(195,70){\vector(1,-2){14}} \put(210,40){\vector(-1,2){14}} 
\put(180,80){\line(3,-2){13}} \put(195,70){\vector(-3,2){13}} 

\put(180,80){\vector(-3,-2){6}} \put(165,70){\vector(3,2){6}} \put(165,70){\line(3,2){15}} 
\put(165,70){\line(-1,-2){15}} \put(150,40){\vector(1,2){14}} 
\put(210,40){\vector(-1,0){58}} \put(150,40){\line(1,0){60}}   
\put(170,55){\scriptsize{123$\,\cap\,$456}}

\end{picture}
\end{center}

On the right hand side we have inverted the edges with a negative face-edge-incidence to illustrate the 
edges cycle. This strange boundary does not define an orientation of the 
face---one part of the boundary surrounds the face in counter-clock-wise direction whereas 
the other part goes clock-wise. 
Both these parts start with a common edge that has two endpoints and end with another common edge with two 
starting points (upper left). 
Anyway, formally this is a valid cycle and the result, in fact, constitutes a complex. 
We suppose that we can take that boundary as it is for our next algorithmic step. 

As we might need an orientation for that step, however, 
we came up so far with an algorithm which iterates the dimensions starting with $1$ (the edges) and attaches 
an arbitrary orientation to each cell. 
Because of the manifold property of each cell we can specify an orientation of $c$ by merely attaching one 
arbitrary sign, say $+1$, to a cell-incidence $R(c,d)$ from $c$ to an already oriented boundary cell $d$. 
This sign then determines the signs of all other incidences of that cell $c$ with its boundary cells. 

Finally the orientation of the $n$\nobreakdash-cells may have to be inverted, because later we want that the 
orientation of each resulting $n$\nobreakdash-cell is the product of the orientations of the intersecting 
$n$\nobreakdash-simplices. 
By now this restricts the validity of the next algorithmic step to full-dimensional complexes, in other words, 
it might only work with $n$\nobreakdash-d complexes in $\bbbr^n$. 

\section{Summing Simplex Intersections}

We simply sum up all these intersection complexes thereby respecting the previously attached signs
to get the resulting intersection complex. If the intersected simplices have different signs then 
the intersection complex will be subtracted from the resulting complex and otherwise added. 
Then the simplex boundary cells that have been introduced by the triangulation will cancel to zero. 
We will show that this at least works with full-dimensional complexes: 

Let $p$ be a point in $\bbbr^n$ and let $\sigma = \simplex{v_0,\dotsc,v_n}$ be an $n$-simplex in $\bbbr^n$. 
Then $p$ is almost never a boundary point and, hence, can be considered either an interior or an exterior point. 
If $p$ is an interior point, the boundary ``wraps around'' $p$ one or several times in some ``direction''. This is 
called the degree (or winding number) of $\sigma$'s boundary with respect to $p$ and is nothing but the orientation 
of $\sigma$ which is non-zero. If $p$ is an exterior point, however, we have winding number $0$. 

But the winding number of a cell's boundary $\partial(c)$ is the same as the sum of the signed winding numbers of 
the simplices' boundaries in the signed decomposition of $c$. 
This is also true for another triangulated $n$-cell $z$ of the other complex. 
Let us denote the winding number of a cell boundary $\partial(c)$ at point $p$ by $w(p,c)$. 
Then for the decomposition $\mu_n(c) = \alpha_1 \sigma_1 + \dotsb + \alpha_k \sigma_k$ into simplices 
$\sigma_i$ we have 
\begin{equation}
   w(p,c) = \alpha_1 w(p,\sigma_1) + \dotsb + \alpha_k w(p,\sigma_k).
\end{equation}

Now $p$ is an interior point of $c\cap z$ iff both winding numbers are non-zero, or, equivalently, iff the product 
of both winding numbers is non-zero. Then we have 
\begin{equation}
  p\in c\cap z \Leftrightarrow w(p,c)\cdot w(p,z) \neq 0
\end{equation}
As we have a boundary for a simplex intersection $\sigma_a\cap\sigma_b$ such that 
$w(p,\sigma_a)\cdot w(p,\sigma_b)=w(p,\sigma_a\cap\sigma_b)$ we can compute $w(p,c)\cdot w(p,z)$ by 
\begin{align}
\nonumber
    w(p,c)\cdot w(p,z) &= (\alpha_1 w(p,\sigma_1) + \dotsb + \alpha_n w(p,\sigma_i))\cdot\\
\nonumber
                       &\phantom{{} = {}}(\beta_1 w(p,\zeta_1) + \dotsb + \beta_m w(p,\zeta_m))\\
                       &= \sum_{i=1}^n\sum_{j=1}^m\alpha_i \beta_j w(p,\sigma_i\cap\zeta_j)
\end{align}
As $w(p,\emptyset)=0$ we only have to sum up the winding numbers of non-empty simplex intersections with coefficients 
$\alpha_i\beta_j$. This is the reason for our above mentioned restriction on the orientation of the simplex intersection. 
Of course, here spatial indexing or a sweeping hyperplane approach would increase efficiency, but we will not 
particularise spatial indexing here.

Then the refinement that stems from the signed decomposition must be undone. Therefore the simplex-simplex 
intersections will be re-composed to get the final result by using the inverse of the two morphisms
$\mu_\mathcal{C}$ and $\mu_\mathcal{K}$. 

Finally, as intersecting possibly non-convex cells may give non-connected resulting cells, it is also 
possible---but not necessary---to identify connected components of cell intersections if this is needed. 

\section{Applications and Outlook}

Polytope complex intersection can be used to compute ``topological relations'', to overlay spatial structures like 
a geological formation and a drilling path or the ``common footprint'' of two geological structures that meet at a fault. 
It can also be used to combine spatial data sets, say, a city model and a GIS data set, into one. 
It is dimension independent and, for example, can also be applied to cut temporal slices out of a 4D space-time 
complex.

So we have shown a relational database schema for polytope complexes of any dimension and an intersection 
algorithm for full-dimensional complexes which seems efficient. 

However, even if this combinatorial decomposition looks quite efficient there is a general complexity problem: 
If we combinatorially triangulate an $n$\nobreakdash-dimensional hypercube we get $n!$ 
simplices. So the number of simplices can grow super-exponentially with the complexity of the triangulated 
polytope if there is no fixed dimension upper bound. 
In practice dimension might be bounded from above by $5$: E.g.\ three spatial dimensions, time, and, maybe, 
version history. Then our triangulation creates $5! = 120$ simplices for a cube. 
In future we therefore want to study polytope complex overlay by using a signed convex cell decomposition 
into more general convex shapes. 
However, we suspect that the above mentioned complexity explosion with unbounded dimension is 
fundamental, yet another curse of dimensionality, and cannot be overcome by any algorithm be it smarter than 
the one presented here or not. 

Additionally, it is worthwhile to further investigate the strange results of ``purely algebraic'' boundary 
operators. First it should be investigated if such strange boundary could simply be tolerated when intersections 
are summed up. One could also search for an alternative to compute intersected cell orientation more 
deterministically than choosing an arbitrary initial ``seed'' incidence sign. 
In particular all this should lead to intersect complexes of different dimension or of lower dimension than 
the embedding space. 

Also, the algorithm computes many simplex-simplex-intersections which later disappear at the re-composition 
phase and, hence, need not be computed. So it might save computation time if these intersections were identified 
in advance and not computed at all.

\section*{Acknowledgements}

This work is funded by the Deutsche Forschungsgemeinschaft (DFG) with research grants BR~2128/12-1 
and BR~3513/3-1. The author thanks Patrick Erik Bradley and Martin Breunig for valuable discussions.


\newpage
\begin{thebibliography}{10}
\providecommand{\url}[1]{{#1}}
\providecommand{\urlprefix}{URL }
\expandafter\ifx\csname urlstyle\endcsname\relax
  \providecommand{\doi}[1]{DOI~\discretionary{}{}{}#1}\else
  \providecommand{\doi}{DOI~\discretionary{}{}{}\begingroup
  \urlstyle{rm}\Url}\fi

\bibitem{Alexandroff:Raeume}
Alexandroff, P.: Diskrete {R}{\"a}ume.
\newblock Matematie{\'{c}}eskij Sbornik \textbf{44}(2), 501--519 (1937).
\newblock \urlprefix\url{http://mi.mathnet.ru/msb5579}

\bibitem{LAPACK:UGuide}
Anderson, E. (ed.): LAPACK users guide, 3. ed. edn.
\newblock Software environments tools. Society for Industrial and Applied
  Mathematics, Philadelphia, Pa. (1999).
\newblock \urlprefix\url{http://www.netlib.org/lapack/lug/}

\bibitem{BradleyPaul:DTop}
Bradley, P.E., Paul, N.: {Using the Relational Model to Capture Topological
  Information of Spaces}.
\newblock The Computer Journal \textbf{53}(1), 69--89 (2010).
\newblock \doi{10.1093/comjnl/bxn054}

\bibitem{BuelerEtAl:VolumeComputation}
B{\"u}eler, B., Enge, A., Fukuda, K.: Polytopes - Combinatorics and
  Computation, chap. Exact volume computation for convex polytopes: A practical
  study.
\newblock Birkh{\"a}user (2000)

\bibitem{Bulbul:AHD}
Bulbul, R., Frank, A.: {AHD}: The alternate hierarchical decomposition of
  nonconvex polytopes.
\newblock In: Geoinformatics, 2009 17th International Conference on, pp. 1--6
  (2009).
\newblock \doi{10.1109/GEOINFORMATICS.2009.5293499}

\bibitem{Bulbul:ConvexDecomp}
Bulbul, R., Karimipour, F., Frank, A.U.: A simplex based dimension independent
  approach for convex decomposition of nonconvex polytopes.
\newblock In: B.G. Lees, S.W. Laffan (eds.) 10th International Conference on
  GeoComputation. UNSW, Sydney (2009)

\bibitem{Chazelle:ConvexPartitions}
Chazelle, B.: Convex partitions of polyhedra: A lower bound and worst-case
  optimal algorithm.
\newblock SIAM Journal on Computing pp. 488--507 vol.13 no.3 (1984)

\bibitem{Chazelle:Triangulation}
Chazelle, B.: Triangulating a simple polygon in linear time.
\newblock Discrete and Computational Geometry pp. 485--524 vol.6 (1991)

\bibitem{Chazelle:DecompAlg}
Chazelle, B., Palios, L.: Decomposition algorithms in geometry.
\newblock In: C.~Bajaj (ed.) Algebraic Geometry and its Applications, chap.~27,
  pp. 419--447. Springer-Verlag (1994)

\bibitem{CohenHickey:Triangulation}
Cohen, J., Hickey, T.: Two algorithms for determining volumes of convex
  polyhedra.
\newblock Journal of the ACM \textbf{26}(3), 401--414 (1979)

\bibitem{EilenbergZilber:Product}
Eilenberg, S., Zilber, J.A.: On products of complexes.
\newblock American Journal of Mathematics \textbf{75}(1), 200--204 (1953)

\bibitem{Hatcher:AlgTop}
Hatcher, A.: {Algebraic Topology}.
\newblock Cambridge University Press (2002).
\newblock \urlprefix\url{http://www.math.cornell.edu/~hatcher/}

\bibitem{Khachiyan:hardPolyhedronVertices}
Khachiyan, L., Boros, E., Borys, K., Elbassioni, K., Gurvich, V.: Generating
  all vertices of a polyhedron is hard.
\newblock Discrete Comput Geom \textbf{39}, 174--190 (2008).
\newblock DOI 10.1007/s00454-008-9050-5

\bibitem{Lawrence:PolytopeVolume}
Lawrence, J.: Ppolytope volume computation.
\newblock Mathematics of Computation \textbf{57}(195), 174--190 (1991)

\bibitem{Lefschetz:ComplexIntersection}
Lefschetz, S.: Intersections and transformations of complexes and manifolds.
\newblock Transactions of the American Mathematical Society \textbf{28}(1),
  1--49 (1926).
\newblock \urlprefix\url{http://www.jstor.org/stable/1989171}

\bibitem{Nocedal:Optimization}
Nocedal, J., Wright, S.J.: Numerical optimization, 2. ed. edn.
\newblock Springer series in operation research and financial engineering.
  Springer, New York, NY (2006)

\bibitem{Streilein:3DDataManagement}
Streilein, A.: 3d data management --- relevance for a 3d cadastre, position
  paper 3.
\newblock 2nd International Workshop on 3D Cadastres  (2011)

\end{thebibliography}
\end{document}